\documentclass[twocolumn,reprint,aps,prl,amsmath,superscriptaddress,amssymb]{revtex4-1}
\usepackage[latin9]{inputenc}
\usepackage{amsmath}
\usepackage{cancel}
\usepackage{graphicx}
\usepackage{esint}

\makeatletter
\@ifundefined{textcolor}{}
{%
 \definecolor{BLACK}{gray}{0}
 \definecolor{WHITE}{gray}{1}
 \definecolor{RED}{rgb}{1,0,0}
 \definecolor{GREEN}{rgb}{0,1,0}
 \definecolor{BLUE}{rgb}{0,0,1}
 \definecolor{CYAN}{cmyk}{1,0,0,0}
 \definecolor{MAGENTA}{cmyk}{0,1,0,0}
 \definecolor{YELLOW}{cmyk}{0,0,1,0}
}

\usepackage{braket}

\renewcommand{\vec}{\mathbf}

\renewcommand{\Im}{\mathop{\mathrm{Im}}}
\renewcommand{\Re}{\mathop{\mathrm{Re}}}

\makeatother

\begin{document}

\title{Origin of the DC and AC conductivity anisotropy in iron-based superconductors: scattering rate versus spectral weight effects}

\author{Michael Schütt}

\affiliation{School of Physics and Astronomy, University of Minnesota, Minneapolis
55455, USA}

\author{Jörg Schmalian}

\affiliation{Institute for Theory of Condensed Matter and Institute for Solid
State Physics, Karlsruhe Institute of Technology, 76128 Karlsruhe,
Germany.}

\author{Rafael M. Fernandes}

\affiliation{School of Physics and Astronomy, University of Minnesota, Minneapolis
55455, USA}
\begin{abstract}
To shed light on the mechanism responsible for the in-plane resistivity
anisotropy in the nematic phase of the iron-based superconductors,
we investigate the impact of spin fluctuations on the anisotropic
AC conductivity. On the one hand, the scattering of electrons off
magnetic fluctuations causes an anisotropic scattering rate. On the
other hand, the accompanying Fermi velocity renormalization contributes
to both the plasma frequency and the scattering rate in antagonistic
ways, giving rise to anisotropies in these quantities that exactly
cancel each other in the DC limit. As a result, the effective scattering
rate changes its sign as function of temperature, but the DC conductivity
retains the same sign, in agreement with recent experiments. 
\end{abstract}
\maketitle
In-plane resistivity anisotropy measurements have been employed as
the primary tool to investigate the nematic phase of both cuprate
and iron-based superconductors~\cite{YoichiPRL2002,JiunHawS2010,TanatarPRB2010,DuszaNJP2012,CyrChoiniereArxiv2015}.
In these systems, the onset of electronic nematic order, characterized
by an Ising order parameter $\varphi\neq0$, lowers the point-group
symmetry from tetragonal to orthorhombic, making the two in-plane
$x$ and $y$ directions inequivalent~\cite{EduardoARCMP2010,VojtaAP2009,FernandesNP2014}.
As a result, a non-zero conductivity anisotropy arises, $\Delta\sigma=\sigma_{x}-\sigma_{y}\neq0$
~\cite{SchuettPRL2015}.

In general, the longitudinal DC conductivity along direction $\mu=x,y$
can be expressed in terms of the Drude form $\sigma_{\mu}=\tau_{\mu}\Omega_{p,\mu}^{2}/(4\pi)$,
where $\tau_{\mu}^{-1}$ is the transport scattering rate (not to
be confused with the inverse single-particle lifetime) and $\Omega_{p,\mu}$
is the plasma frequency. Therefore, an anisotropy in the DC conductivity
can arise from an anisotropic scattering rate, which is sensitive
to impurities and to the low-energy excitations of the system, and/or
from an anisotropic Drude weight, which is sensitive to the electronic
structure. In the nematic phase of the iron-based superconductors,
different effects contribute to these quantities. Anisotropic magnetic
fluctuations triggered by nematic order~\cite{FernandesSSaT2012,LuS2014}
give rise to an anisotropy in the inelastic scattering rate~\cite{FernandesPRL2011,LiangPRL2012,BreitkreizPRB2014},
whereas the dressing of an impurity potential by magnetic correlations
promotes an anisotropy in the elastic scattering rate~\cite{AllanNP2013,GastiasoroPRL2014}.
Conversely, the distortion of the Fermi surface caused by the ferro-orbital
order triggered at the nematic transition affect the plasma frequency~\cite{ChenPRB2010,WeichengPRB2011,ValenzuelaPRL2010,YinNP2011}.
Disentangling these contributions would provide important insight
into the dominant processes responsible for the nematic instability.

At first sight, a natural way to disentangle $\tau_{\mu}^{-1}$ and
$\Omega_{p,\mu}$ is the AC conductivity, $\sigma_{\mu}\left(\omega\right)$.
Indeed, recent measurements of $\sigma_{\mu}\left(\omega\right)$
in detwinned BaFe$_{2}$As$_{2}$ reported a larger anisotropy in
the plasma frequency than in the scattering rate~\cite{DuszaNJP2012,MirriPRL2015},
which was interpreted as evidence for electronic-structure induced
anisotropy. However, as we show in this paper, these two quantities
are unavoidably entangled. This general result follows directly from
the memory function formalism, which is valid even in the absence
of quasi-particles, and yields the following form for the AC conductivity~\cite{GotzePRB1972}:

\begin{equation}
\sigma_{\mu}(\omega)=\frac{\Omega_{p,\mu}^{2}}{4\pi}\frac{1}{\tau_{\mu}^{-1}(\omega)-i\omega\left[1+\lambda_{\mu}(\omega)\right]}.\label{EqGeneralMemoryFunction}
\end{equation}

\begin{figure}
\includegraphics[width=0.49\columnwidth]{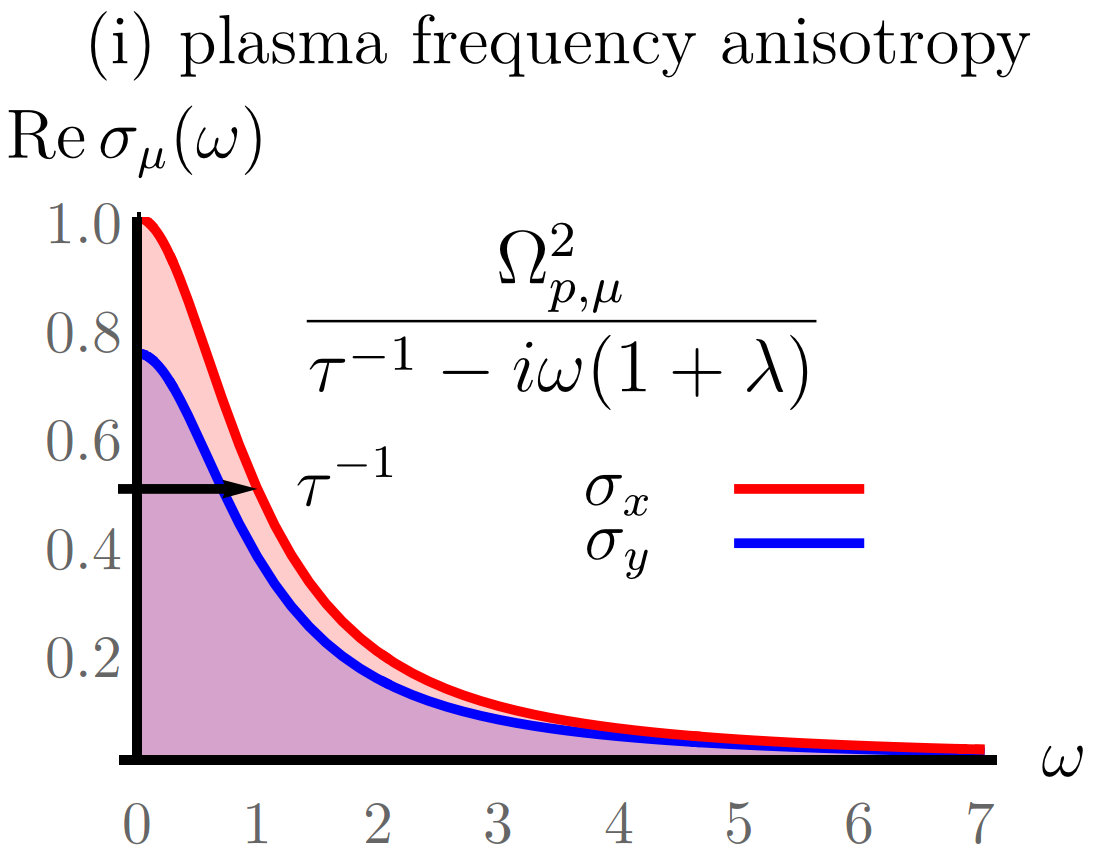}
\includegraphics[width=0.49\columnwidth]{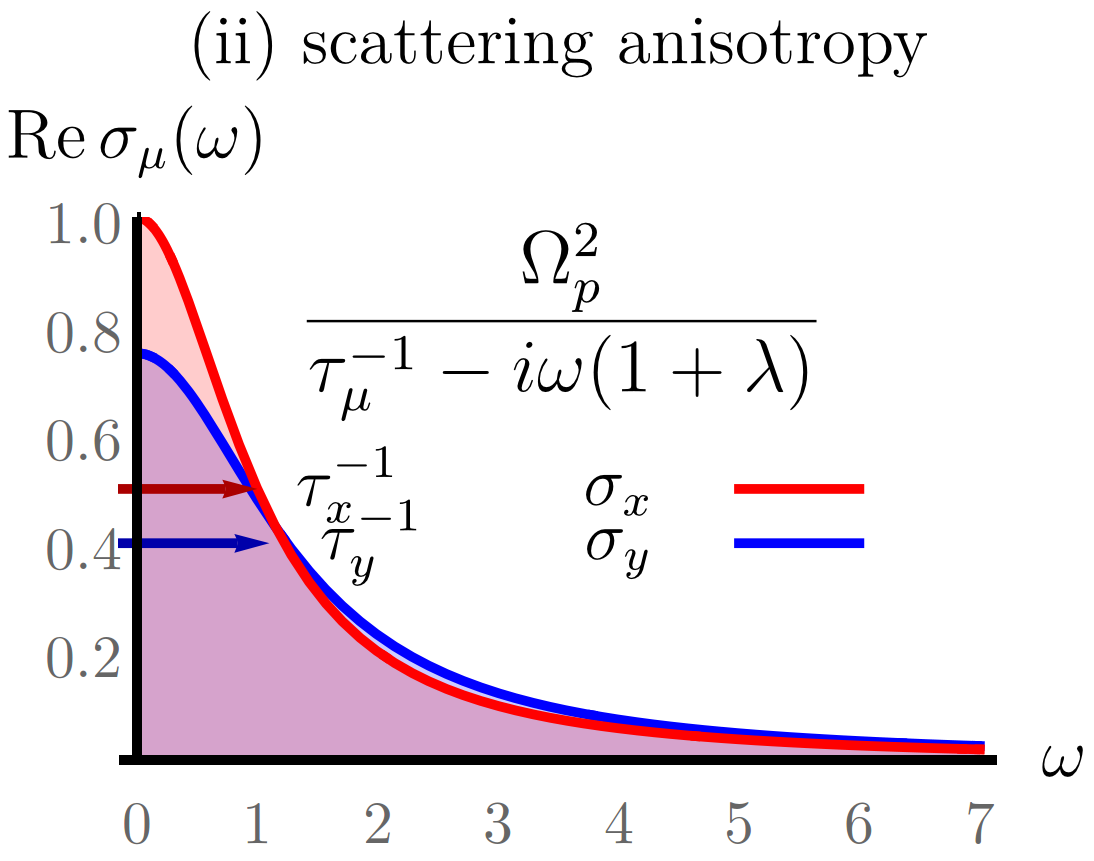}\\
 \includegraphics[width=0.49\columnwidth]{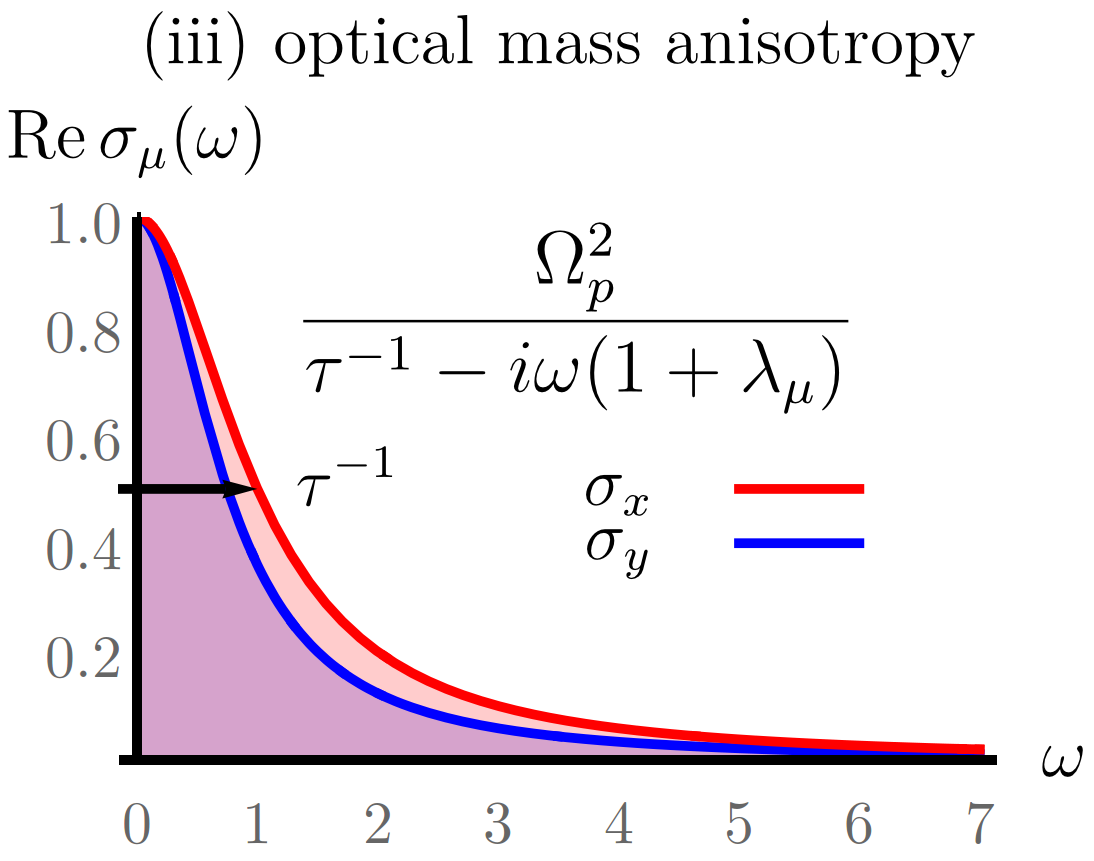}
\includegraphics[width=0.49\columnwidth]{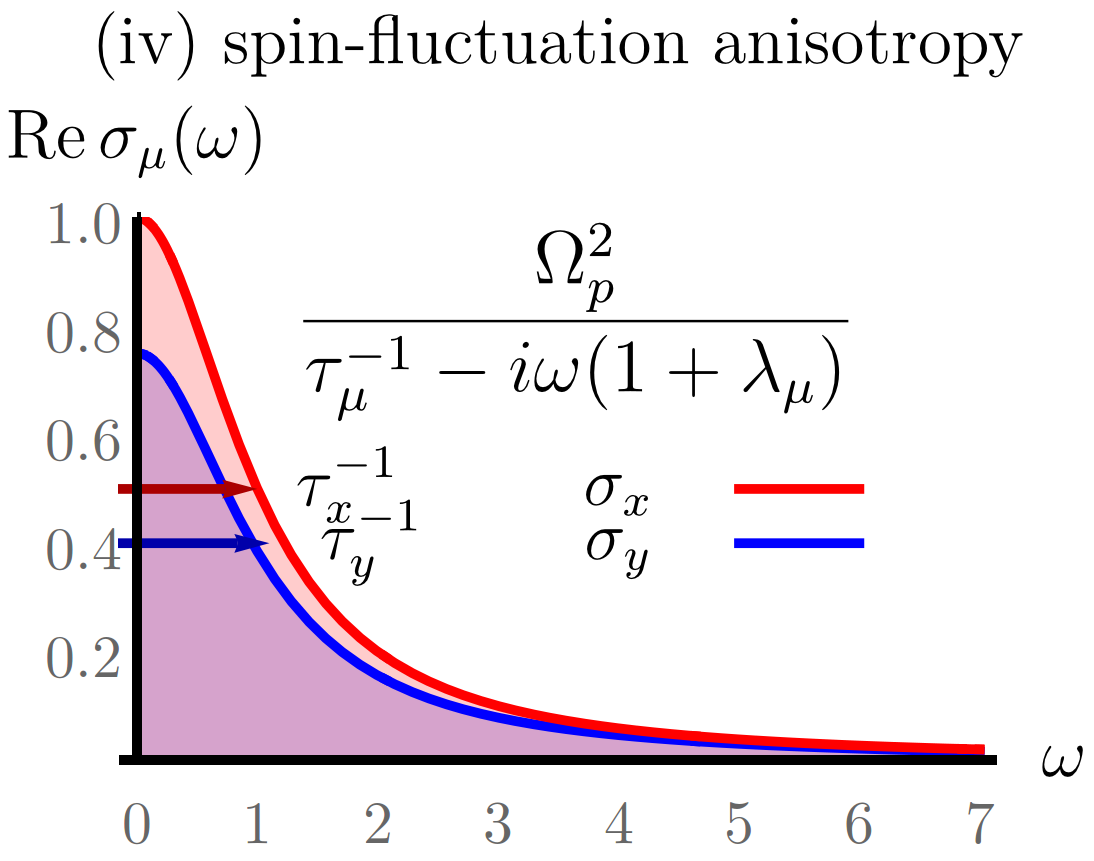}
\protect\protect\protect\caption{AC conductivity $\sigma_{\mu}\left(\omega\right)$ of Eq. (\ref{EqGeneralMemoryFunction})
as function of frequency $\omega$ for both $x$ and $y$ directions.
In case (i)\textbf{,} only the plasma frequency is anisotropic ($\Omega_{p,y}^{2}=(3/4)\Omega_{p,x}^{2}$),
while in (ii) and (iii) only the scattering rate ($\tau_{y}^{-1}=(4/3)\tau_{x}^{-1}$)
and the optical mass ($\lambda_{y}=(4/3)\lambda_{x}$) are anisotropic,
respectively. Panel (iv) shows the case in which both the scattering
rate and the optical mass are anisotropic, yielding AC and DC conductivities
very similar to those in panel (i). The causality properties of the
AC conductivity do not allow cases (ii) and (iii) to exist, i.e. both
$\tau_{\mu}^{-1}$ and $\lambda_{\mu}$ must be present. }

\label{FigCompareEffectOfRen} 
\end{figure}

The main point is that, besides $\Omega_{p,\mu}$ and $\tau_{\mu}^{-1}$,
the AC conductivity depends additionally on the optical mass enhancement
$\lambda_{\mu}$~\cite{BasovRMP2005}. While $\lambda_{\mu}$ does
not contribute to the DC conductivity $\sigma_{\mu}(\omega\rightarrow0)$,
it does modify the effective plasma frequency and the effective scattering
rate extracted from a typical AC conductivity measurement, yielding
$\tilde{\Omega}_{p,\mu}^{2}=\Omega_{p,\mu}^{2}/\left(1+\lambda_{\mu}\right)$
and $\tilde{\tau}_{\mu}^{-1}=\tau_{\mu}^{-1}/\left(1+\lambda_{\mu}\right)$.
To illustrate the importance of this effect for the intepretation
of the experimental results of Ref.~\cite{MirriPRL2015}, in Fig.
\ref{FigCompareEffectOfRen} we plot $\mathrm{Re}\left[\sigma_{\mu}\left(\omega\right)\right]$
in Eq. (\ref{EqGeneralMemoryFunction}) considering four cases: (i)
anisotropy only in the plasma frequency ($\Omega_{p,y}<\Omega_{p,x}$);
(ii) anisotropy only in the scattering rate ($\tau_{y}^{-1}>\tau_{x}^{-1}$);
(iii) anisotropy only in the optical mass ($\lambda_{y}>\lambda_{x}$);
(iv) anisotropy in both scattering rate and optical mass ($\tau_{y}^{-1}>\tau_{x}^{-1}$
and $\lambda_{y}>\lambda_{x}$). The system in cases (i) and (ii)
have very similar DC conductivities, but rather different AC conductivities
in the intermediate frequency range. In the presence of optical mass
anisotropy only, case (iii), the system does not display a DC conductivity
anisotropy, but in the intermediate frequency range, the AC conductivity
is similar to that of case (i). The key point is that cases (ii) and
(iii) are not allowed due to the causality properties of the AC conductivity,
which require \emph{both} $\tau_{\mu}^{-1}$ and $\lambda_{\mu}$
to be present. After combining these two effects, case (iv), the system
displays DC and AC conductivities very similar to case (i). Thus,
the AC conductivity anisotropy observed in Ref.~\cite{MirriPRL2015}
is in equally consistent with either case (i) or (iv), which have
very different physical origins -- electronic-structure anisotropy
and scattering rate anisotropy, respectively.

\begin{figure}
\includegraphics[width=0.8\columnwidth]{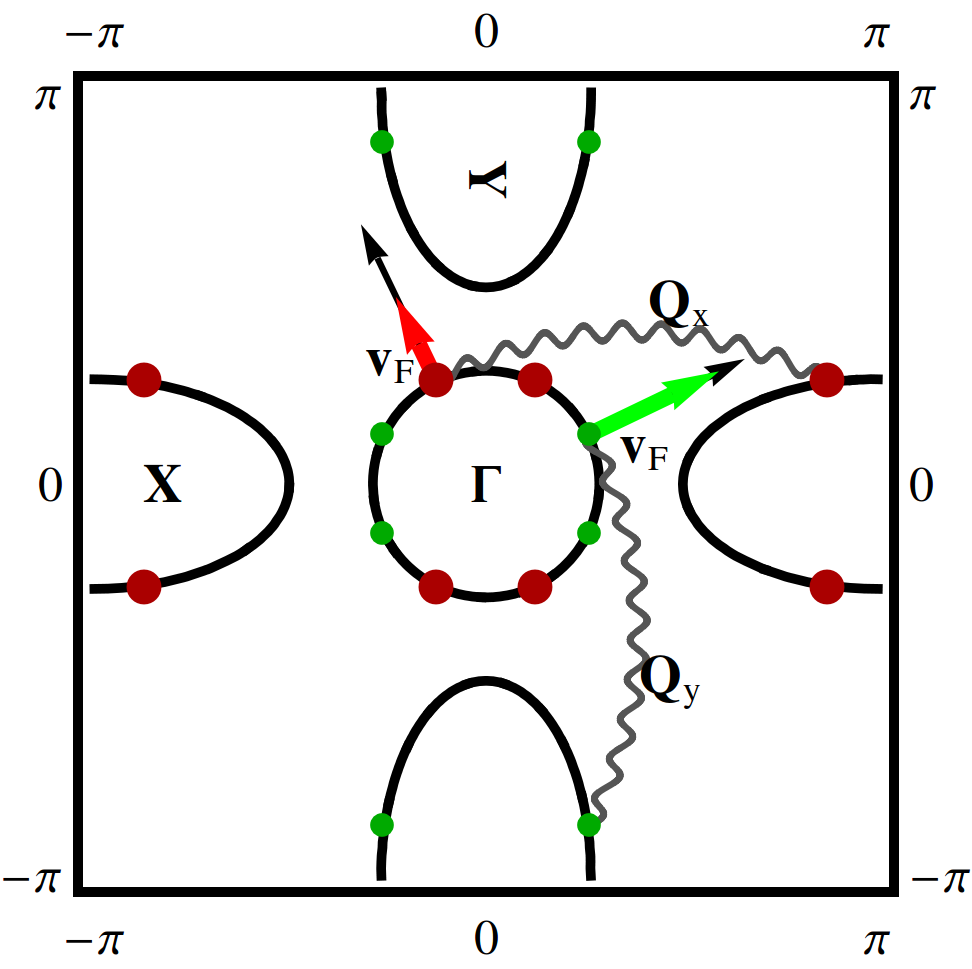}
\protect\protect\protect\caption{Illustration of the Fermi surface, consisting of hole-like (at $\Gamma$)
and electron-like pockets (at $X$ and $Y$), and the anisotropic
processes promoted by spin fluctuations. On the one hand, scattering
off of spin fluctuations is stronger for the hot spots exchanging
$\mathbf{Q}_{X}=\left(\pi,0\right)$ (red dots) than $\mathbf{Q}_{Y}=(0,\pi)$
(green dots) fluctuations. On the other hand, the renormalized Fermi
velocity suppression (or, equivalently, mass renormalization) caused
by the exchange of spin fluctuations (arrows) is smaller at the red
hot spots than at the green hot spots.}

\label{FigBandStructureAndRenormalizationIllustration} 
\end{figure}

In this paper we compute the AC conductivity of a multi-band model
for the iron pnictides in which the electrons interact with spin fluctuations
-- which become anisotropic in the nematic phase \cite{FernandesSSaT2012,LuS2014}.
While this interaction does not promote anisotropy in the bare plasma
frequency, it causes anisotropies simultaneously in both the scattering
rate, $\Delta\tau^{-1}\equiv\tau_{x}^{-1}-\tau_{y}^{-1}\neq0$, and
in the optical mass, $\Delta\lambda\equiv\lambda_{x}-\lambda_{y}\neq0$.
Physically, the first effect arises from real collisions of electrons
and magnetic fluctuations, whereas the latter effect stems from the
reduction of the electronic Fermi velocity (or, equivalently, the
enhancement of the effective electron mass) promoted by the exchange
of virtual spin fluctuations (see Fig. \ref{FigBandStructureAndRenormalizationIllustration}). 

More importantly, our microscopic calculations reveal that while the
overall signs of both quantities, $\Delta\tau^{-1}$ and $\Delta\lambda$,
are determined by the same prefactor that depends on the geometry
of the Fermi surface \cite{FernandesPRL2011}, the relative sign of
$\Delta\tau^{-1}$ and $\Delta\lambda$ is always positive, similar
to case (iv) in Fig. \ref{FigCompareEffectOfRen}. Consequently, the
\emph{effective} scattering rate anisotropy $\Delta\tilde{\tau}^{-1}=\Delta\tau^{-1}-\tau_{0}^{-1}\Delta\lambda$
is reduced with respect to its bare value, while the \emph{effective}
plasma frequency anisotropy $\Delta\tilde{\Omega}_{p}^{2}=-\Omega_{p,0}^{2}\Delta\lambda$
is enhanced (here $\tau_{0}^{-1}$ and $\Omega_{p,0}$ are the bare
isotropic scattering rate and plasma frequency). Interestingly, because
collisions are suppressed at low temperatures, $\Delta\tau^{-1}$
decreases as the temperature is lowered. In contrast, $\Delta\lambda$
remains finite as $T\rightarrow0$, since it is proportional to the
electronic mass renormalization. Consequently, one generally expects
a sign-change of $\Delta\tilde{\tau}^{-1}$ as function of temperature,
despite the fact that $\Delta\sigma\left(\omega\rightarrow0\right)\propto\Delta\tau^{-1}$
retains the same sign. These behaviors agree with recent AC conductivity
measurements in detwinned BaFe$_{2}$As$_{2}$ \cite{DuszaNJP2012,MirriPRL2015},
revealing the importance of anisotropic spin fluctuations to describe
the nematic state of the iron pnictides.

Our starting point is the minimal three band model shown in Fig.~\ref{FigBandStructureAndRenormalizationIllustration}~\cite{FernandesKnollePRB2012},
and previously employed to investigate the DC conductivity anisotropy
due to the scattering by spin fluctuations \cite{FernandesPRL2011}.
This model has a hole-like circular pocket centered at the $\Gamma=\left(0,0\right)$
point of the Fe-square lattice Brillouin zone, and two elliptical
electron pockets centered at $X=\left(\pi,0\right)$ and $Y=\left(0,\pi\right)$.
Hereafter, for convenience, these bands are labeled $\beta=0$, $\beta=1$,
and $\beta=-1$, respectively. We also include in the model point-like
impurities, giving rise to the isotropic band-independent elastic
scattering rate $\tau_{0}^{-1}$.

To make the symmetry properties of the AC conductivity more evident,
we consider the band-resolved conductivity $\sigma_{\mu}^{\beta}$
for band $\beta$ in direction $\mu$ such that, $\sigma_{\mu}=\sum_{\beta}\sigma_{\mu}^{\beta}$.
Without interactions, we have $\sigma_{0,\mu}^{\beta}=\frac{1}{4\pi}\left(\Omega_{p,\mu}^{\beta}\right)^{2}/(\tau_{0}^{-1}-i\omega)$,
where the subscript $0$ denotes the non-interacting system and $\Omega_{p,\mu}^{\beta}=\sqrt{\frac{2e^{2}N_{F}^{\beta}}{\hbar}}v_{F}^{\beta}$,
with density of states $N_{F}^{\beta}$ and averaged Fermi velocity
$v_{F}^{\beta}$. The tetragonal symmetry of the system implies $\Omega_{p,x}^{\beta}=\Omega_{p,y}^{-\beta}$,
i.e. $\sigma_{0,x}^{\beta}=\sigma_{0,y}^{-\beta}$, yielding $\Delta\sigma_{0}\equiv\sigma_{0,x}-\sigma_{0,y}=0$,
as expected for a tetragonal system.

The contribution arising from the interaction with spin fluctuations
is conveniently expressed in terms of the memory function $M_{\mu}^{\beta}\left(\omega\right)$,
defined such that $\sigma_{\mu}^{\beta}=\frac{1}{4\pi}\left(\Omega_{p,\mu}^{\beta}\right)^{2}/\left[\tau_{0}^{-1}-i\omega+M_{\mu}^{\beta}\left(\omega\right)\right]$
\cite{GotzePRB1972}. Hence, while $\mathrm{Re}\left(M_{\mu}^{\beta}\right)$
renormalizes the scattering rate, $-\mathrm{Im}\left(M_{\mu}^{\beta}\right)/\omega$
renormalizes the optical mass. The fact that these two quantities
are related by Kramers-Kronig relations implies that an anisotropy
in the scattering rate must be accompanied by an anisotropy in the
optical mass, as stated in the introduction. In this framework, calculating
the band-resolved AC conductivity anisotropy, $\Delta\sigma^{\beta}\equiv\Delta\sigma_{x}^{\beta}-\Delta\sigma_{y}^{\beta}$,
is equivalent to calculating the anisotropic memory function $\Delta M^{\beta}=M_{x}^{\beta}-M_{y}^{-\beta}$,
since $\Delta\sigma^{\beta}=-\left(\frac{\sigma_{0,x}^{\beta}}{\tau_{0}^{-1}-i\omega}\right)\Delta M^{\beta}$.
Consequently, expansion of $\Delta M^{\beta}$ for small frequencies
yields the anisotropic scattering rate and optical mass, $\Delta M^{\beta}=\left(\Delta\tau^{\beta}\right)^{-1}-i\omega\Delta\lambda^{\beta}$.

\begin{figure}
\includegraphics[width=0.45\columnwidth]{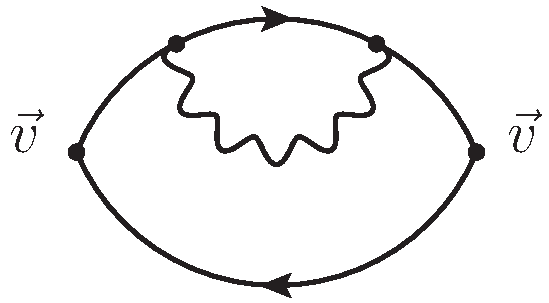}
\includegraphics[width=0.45\columnwidth]{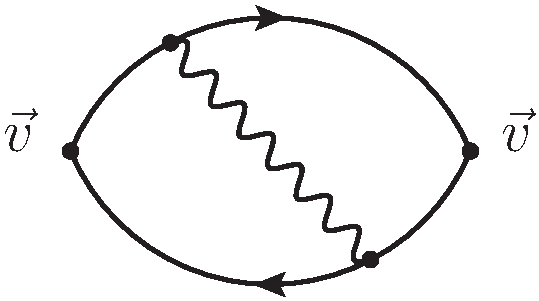}
\protect\protect\protect\caption{Interaction corrections to the optical conductivity diagrams: self-energy
$\Phi^{\mathrm{self}}$ (left) and vertex corrections $\Phi^{\mathrm{vert}}$
(right). Solid lines refer to the electronic propagator, wavy lines
denote the spin fluctuation propagator, and $\vec{\mathbf{v}}$ is
the velocity vertex. }

\label{FigDiagrams} 
\end{figure}

To leading order in the interaction parameter $g$ between the electrons
and the spin fluctuations, the memory function is given by the two
Feynman diagrams depicted in Fig.~\ref{FigDiagrams}~\cite{SyzranovPRL2012},
where solid lines denote the electronic Green's function $\mathcal{G}_{k}^{\beta}=(i\tilde{\omega}_{n}-\epsilon_{\mathbf{k}}^{\beta})^{-1}$
and the wavy lines denote the spin fluctuation dynamic susceptibility
$\chi_{k}$. Here, $k=(i\omega_{n},\vec{k})$ denotes both momentum
$\mathbf{k}$ and Matsubara frequency $\omega_{n}$, $\epsilon_{\mathbf{k}}^{\beta}\approx\mathbf{v}_{F}^{\beta}\cdot\mathbf{k}$
is the linearized dispersion of band $\beta$, and $\tilde{\omega}_{n}=\omega_{n}+\mathrm{sgn}(\omega_{n})/(2\tau_{0})$
is introduced in the Green's function to incorporate the effect of
impurity scattering within the Born approximation. In particular,
the memory function is given by the combination of diagrams $\left(\frac{\sigma_{0,\mu}^{\beta}}{\tau_{0}^{-1}-i\omega}\right)M_{\mu}^{\beta}=\frac{e^{2}}{i\omega}(2\Phi_{\mu}^{\mathrm{self},\beta}+\Phi_{\mu}^{\mathrm{vert},\beta})$
with: 
\begin{align}
\Phi_{\mu,p}^{\mathrm{self},\beta} & =g^{2}\sum_{\substack{\beta'}
}\int_{k,k'}\chi_{k-k'}^{\left(\beta\beta'\right)}\mathcal{G}_{k}^{\beta}v_{\mu,k}^{\beta}\mathcal{G}_{k+p}^{\beta}\mathcal{G}_{k'+p}^{\beta'}\mathcal{G}_{k+p}^{\beta}v_{\mu,k+p}^{\beta},\label{EqDiagrammaticInteractionCorrection}\\
\Phi_{\mu,p}^{\mathrm{vert},\beta} & =g^{2}\sum_{\substack{\beta'}
}\int_{k,k'}\chi_{k-k'}^{\left(\beta\beta'\right)}\mathcal{G}_{k}^{\beta}v_{\mu,k}^{\beta}\mathcal{G}_{k+p}^{\lambda}\mathcal{G}_{k'}^{\beta'}v_{\mu,k'+p}^{\beta'}\mathcal{G}_{k'+p}^{\beta'},
\end{align}
where $v_{\mu,k}^{\beta}$ is the velocity, $p=(i\Omega_{n},0)$,
and $\int_{k}=T\sum_{\omega_{n}}\int_{\mathrm{BZ}}\mathrm{d}^{2}k/(2\pi)^{2}$.
As shown by inelastic neutron scattering data~\cite{InosovNP2010},
the magnetic susceptibility is peaked at the ordering vectors $\mathbf{Q}_{X}=\left(\pi,0\right)$
and $\mathbf{Q}_{Y}=\left(0,\pi\right)$. Therefore, the only terms
that contribute to the sums above are $\chi_{k}^{\left(10\right)}\equiv\chi_{X,k}$
and $\chi_{k}^{\left(-10\right)}\equiv\chi_{Y,k}$. At low energies
and in the tetragonal phase, these susceptibilities are equal and
described by $\chi_{j,k}^{-1}=\chi_{0}^{-1}\left(\xi^{-2}+|\vec{k}-\vec{Q}_{j}|^{2}+\left|\omega_{n}\right|/\gamma\right)$,
where $\chi_{0}^{-1}$ is the magnetic energy scale, $\xi$ is the
correlation length (in units of the lattice constant), and $\gamma$
is the Landau damping. Indeed, this form has been widely used to fit
the neutron data in pnictides~\cite{InosovNP2010,DialloPRB2010,TuckerPRB2012}.
In the nematic phase, the susceptibilities become different, since
magnetic fluctuations become stronger along either $\mathbf{Q}_{X}$
or $\mathbf{Q}_{Y}$. Specifically, the correlation lengths are renormalized
by the nematic order parameter $\varphi$, yielding $\chi_{j,k}^{-1}=\chi_{j,k}^{-1}\mp\chi_{0}^{-1}\xi^{-2}\varphi$~\cite{FernandesSSaT2012}.

Such an anisotropy in the spin fluctuation spectrum, which is indeed
observed experimentally~\cite{LuS2014}, is responsible for the anisotropy
in the memory function. Consequently, it is useful to expand $\Delta M^{\beta}$
for small $\varphi$. In contrast to the isotropic contribution to
the memory function~\cite{AbanovAP2003}, the behavior of $\Delta M^{\beta}$
is dominated by the hot spots of the Fermi surface -- i.e. points
of the Fermi surface connected by the magnetic ordering vectors $\mathbf{Q}_{X}=\left(\pi,0\right)$
and $\mathbf{Q}_{Y}=\left(0,\pi\right)$, $\epsilon_{\mathbf{k}}^{\beta}=\epsilon_{\mathbf{k}+\mathbf{Q}_{j}}^{\beta'}$
(see Fig. \ref{FigBandStructureAndRenormalizationIllustration}).
Focusing on this contribution, we find the analytical expression:

\begin{figure}
\includegraphics[width=0.8\columnwidth]{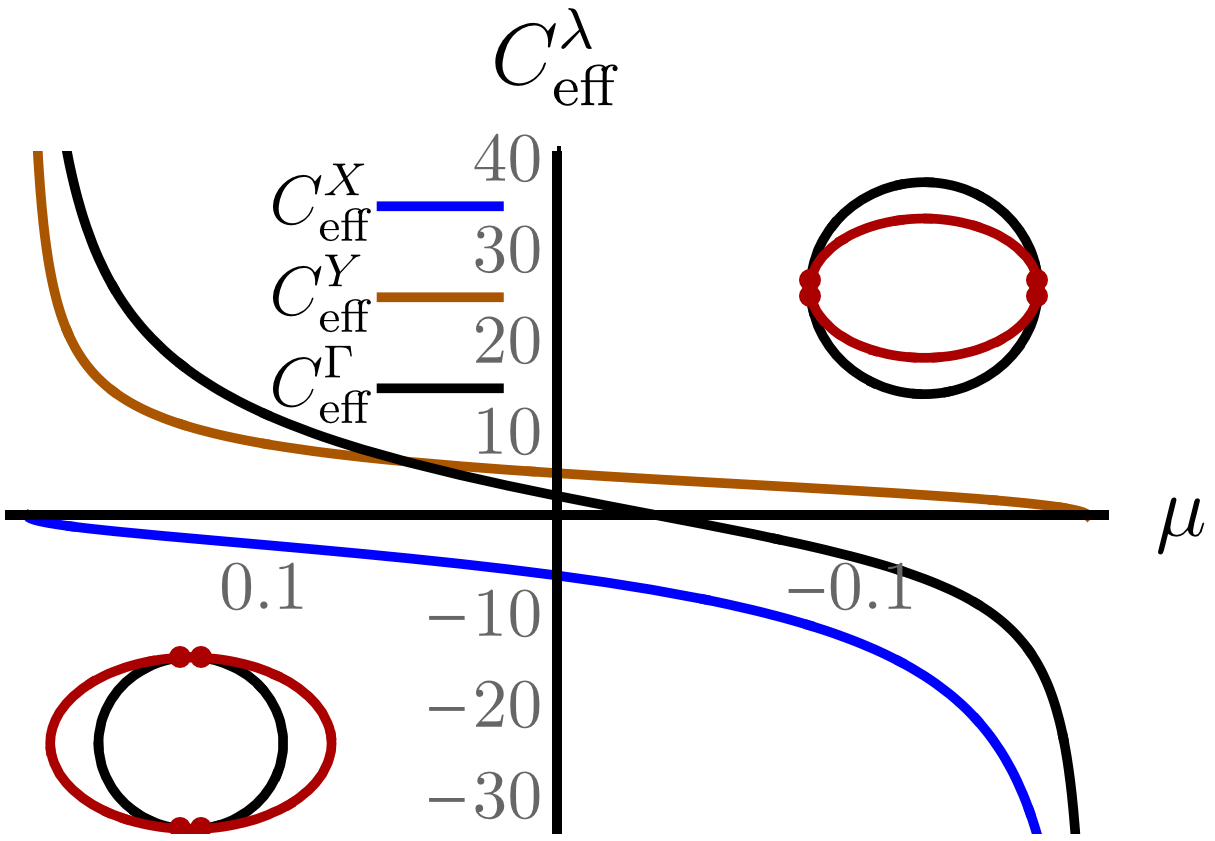} \protect\protect\protect\caption{The coefficient $C_{\mathrm{eff}}^{\beta}$, as function of the chemical
potential $\mu$ (in units of the Fermi energy $\epsilon_{0}$), for
a circular hole pocket at the $\Gamma$ point and elliptical electron
pockets centered at $X$ and $Y$. The ellipticity here is set to
$\delta=0.35$.}

\label{FigTheCoefficient} 
\end{figure}

\begin{equation}
\Delta M^{\beta}(\omega)=-\varphi\tilde{g}^{2}C_{\mathrm{eff}}^{\beta}\omega\,\mathcal{K}\left(\frac{\xi^{-2}}{2\pi}\frac{\gamma}{T},\frac{\omega}{\xi^{-2}\gamma}\right).\label{EqAnisotropicMemoryFunction}
\end{equation}
where we defined the dimensionless coupling constant $\tilde{g}{}^{2}=g^{2}\chi_{0}\nu_{F}^{\beta}$
and the complex function $\mathcal{K}\left(s,t\right)$: 
\begin{multline}
\mathcal{K}(s,t)=-\frac{1}{t}\left[1-\frac{1}{2s}+\left(1+\frac{i}{t}\right)\times\right.\\
\left.\left(\frac{1}{s}+\psi(s)-\psi(1+s-ist)\right)\right],\label{EqKfunctionExplicit}
\end{multline}
with $\psi$ denoting the digamma function. Eq.~\eqref{EqAnisotropicMemoryFunction}
naturally separates the contributions arising from the Fermi surface
geometry, enconded in $C_{\mathrm{eff}}^{\beta}$ (see supplementary~\footnote{See Supplemental Material [url], which includes Refs.\cite{SchuettPRL2015,JohnstonAP2010}} material for the full expression), and the contributions arising from
the spin dynamics, encoded in $\mathcal{K}\left(s,t\right)$ via the
two dimensionless parameters $s\equiv\xi^{-2}\gamma/(2\pi T)$ and
$t=\omega/\left(\xi^{-2}\gamma\right)$. While $s$ is proportional
to the ratio between the spin correlation length $\xi$ and the length
scale of thermal spin fluctuations $\xi_{T}^{2}\equiv\gamma/T$, $t$
depends explicitly on the frequency. Because we are interested in
the interaction-induced corrections to the standard Drude formula,
hereafter we take the limit $t\ll1$. Terms beyond this approximation
are particularly important near quantum critical points, where $\xi\rightarrow\infty$,
and at frequencies larger than the scale set by the isotropic scattering
rate $\tau_{0}^{-1}$, which is of the order of $300$ meV in BaFe$_{2}$As$_{2}$~(see
supplementary material and also \cite{JohnstonAP2010}). Although
the study of these contributions is beyond the scope of this paper,
we note that for $\omega\gg\tau_{0}^{-1}$ they give rise to a slower
decay of $\mathrm{Re}[\sigma\left(\omega\right)]$ than the standard
$\omega^{-2}$ Drude behavior~%
\footnote{A detailed study of the isotropic AC conductivity in this limit will
be reported elsewhere%
}.

In terms of the function $\mathcal{K}\left(s,t\right)$, the anisotropies
in the bare scattering rate and in the optical mass are given by $\left(\Delta\tau^{\beta}\right)^{-1}=-\varphi\tilde{g}^{2}C_{\mathrm{eff}}^{\beta}\omega\Re\mathcal{K}\left(s,t\rightarrow0\right)$
and $\Delta\lambda^{\beta}=\varphi\tilde{g}^{2}C_{\mathrm{eff}}^{\beta}\Im\mathcal{K}\left(s,0\right)$,
yielding the effective scattering rate and plasma frequency anisotropies:

\begin{align}
\left(\Delta\tilde{\tau}^{\beta}\right)^{-1} & =-\varphi g'^{2}C_{\mathrm{eff}}^{\beta}\left[\omega\Re\mathcal{K}\left(s,t\rightarrow0\right)+\tau_{0}^{-1}\Im\mathcal{K}\left(s,0\right)\right]\nonumber \\
\left(\Delta\tilde{\Omega}_{p}^{\beta}\right)^{2} & =-\varphi\tilde{g}^{2}C_{\mathrm{eff}}^{\beta}\left(\Omega_{p,x}^{\beta}\right)^{2}\Im\mathcal{K}\left(s,0\right)\label{eq_effective_finals}
\end{align}

As it can be confirmed by explicit evaluation of Eq. (\ref{EqKfunctionExplicit}),
the analytical properties of the complex function $\mathcal{K}\left(s,t\right)$
enforce its real part to be positive and its imaginary part to be
negative. This is because the former arises from the collision of
electrons by spin fluctuations -- the same process that causes a finite
electronic lifetime via the imaginary part of the self-energy -- whereas
the latter arises from the suppression of the electronic Fermi velocity
-- the same process that enhances the electronic mass via the real
part of the self-energy (see Fig. \ref{FigCompareEffectOfRen}). Consequently,
the two contributions to $\left(\Delta\tilde{\tau}^{\beta}\right)^{-1}$
in Eq. (\ref{eq_effective_finals}) have opposite signs, resulting
in a suppression of the effective scattering rate compared to the
bare scattering rate $\left(\Delta\tau^{\beta}\right)^{-1}$. Furthermore,
because of their different physical origins -- inelastic collision
\emph{versus} Fermi velocity renormalization -- the two contributions
to the effective scattering rate $\left(\Delta\tilde{\tau}^{\beta}\right)^{-1}$
display rather different temperature dependencies. Using Eq. (\ref{EqKfunctionExplicit}),
we find that at high temperatures ($T\gg\gamma\xi^{-2}$) the behaviors
$\omega\Re\mathcal{K}\left(s,0^{+}\right)\propto T$ and $\Im\mathcal{K}\left(s,0\right)\propto-\frac{1}{T}$,
whereas at low temperatures ($T\ll\gamma\xi^{-2}$) we have $\omega\Re\mathcal{K}\left(s,0^{+}\right)\propto T^{2}$
and $\Im\mathcal{K}\left(s,0\right)=-\frac{1}{2}$. Thus, while $\Re\mathcal{K}>0$
dominates the high-temperature regime, $\Im\mathcal{K}<0$ governs
the low-temperature regime. This can be physically understood from
the fact that $\Re\mathcal{K}$ arises from the physical collision
between electrons and spin fluctuations, which are completely suppressed
at $T=0$, whereas $\Im\mathcal{K}$ arises from the suppression of
the Fermi velocity, which persists down to $T=0$.

Therefore, one expects that as temperature is lowered, the anisotropy
of the effective scattering rate changes sign. Using characteristic
values for BaFe$_{2}$As$_{2}$ (see supplementary material and also
\cite{JohnstonAP2010}), $\tau_{0}^{-1}\sim300$ meV from the residual
resistivity and $\gamma\sim100$ meV , $\xi\sim5a$ from neutron scattering~\cite{TuckerPRB2012},
we find that the two contributions become comparable at the temperature
scale $T^{*}\sim100$ K. Interestingly, recent optical conductivity
data in this compound \cite{MirriPRL2015} find such a behavior, with
$\left(\Delta\tilde{\tau}^{\beta}\right)^{-1}$ changing sign below
the nematic transition at $T_{\mathrm{nem}}\sim150$ K. Although these
compounds display also long-range magnetic order, at these temperatures
the resulting reconstruction of the Fermi surface is incipient~\cite{LiuPRB2011},
suggesting that the mechanism discussed here could be at play.

We emphasize that the sign change in the effective scattering rate
$\left(\Delta\tilde{\tau}^{\beta}\right)^{-1}$ does not cause a sign
change in the DC conductivity anisotropy -- also in agreement with
the experiments. Indeed, as it is clear from Eq. (\ref{EqGeneralMemoryFunction}),
the DC conductivity anisotropy depends only on the bare scattering
rate $\left(\Delta\tau^{\beta}\right)^{-1}$, which in turn is solely
determined by $\Re\mathcal{K}$, $\Delta\sigma\left(\omega=0\right)=-\frac{1}{4\pi}\sum_{\beta}\left(\Omega_{0,x}^{\beta}\right)^{2}\tau_{0}^{2}\left(\Delta\tau^{\beta}\right)^{-1}$.
The main consequence of the reduction of the effective scattering
rate $\left(\Delta\tilde{\tau}^{\beta}\right)^{-1}$ is an accompanying
enhancement of the anisotropic Drude spectral weight $\Delta\mathrm{SW}\equiv\int_{0}^{\infty}\Delta\sigma\left(\omega\right)d\omega$,
since $\Delta\mathrm{SW}=\frac{1}{8}\sum_{\beta}\left(\Delta\tilde{\Omega}_{p}^{\beta}\right)^{2}$
depends only on $\Im\mathcal{K}$, as shown in Eq. (\ref{eq_effective_finals}).
Physically, this means that any suppression of the effective scattering
rate is compensated by an enhancement of the effective Drude weight,
keeping the DC anisotropy the same.

The global sign of $\Delta\sigma\left(\omega=0\right)$ and $\Delta\mathrm{SW}$
depend on the same parameters $C_{\mathrm{eff}}^{\beta}$ via Eq.
(\ref{eq_effective_finals}), which are determined by the Fermi surface
geometry. We calculate them explicitly in Fig. \ref{FigTheCoefficient}
for a toy model in which the hole pocket is a circle, $\epsilon_{\Gamma}=\epsilon_{0}-\frac{\vec{p}^{2}}{2m}$,
whereas the electron pockets are ellipses, $\epsilon_{X/Y}=\frac{p_{x}^{2}}{2m(1\pm\delta)}+\frac{p_{y}^{2}}{2m(1\mp\delta)}-\epsilon_{0}$~\cite{FernandesPRL2011,EreminPRB2010}.
By fixing the ellipticity $\delta$, we find that in general the weighted
sum of $C_{\mathrm{eff}}^{\beta}$ is positive for electron-doped
compounds ($\mu>0$) and negative for hole-doped compounds ($\mu<0$).
Consequently, because $\varphi>0$ for a detwinned sample with tensile
strain applied along the $x$ direction~\cite{LuS2014}, we find
$\Delta\sigma>0$ and $\Delta\mathrm{SW}>0$ for electron-doped compounds,
and $\Delta\sigma<0$ and $\Delta\mathrm{SW}<0$ for hole-doped compounds.
This agrees with previous theoretical calculations using the Boltzmann
equation instead of the diagrammatic approach used here~\cite{FernandesPRL2011,BreitkreizPRB2014},
as well as with experiments~\cite{BlombergNC2013,KuoArxiv2015}.

In summary, we studied the impact of anisotropic spin fluctuations
on the optical conductivity anisotropy of the nematic phase of iron-based
superconductors. Our main result is that, in this case, while the
DC conductivity anisotropy is determined solely by the collision of
electrons and spin fluctuations, the electronic Fermi velocity renormalization
induced by spin fluctuations causes opposite changes in the effective
scattering rate and plasma frequencies anisotropies that exactly compensate
each other in the DC limit. Our results qualitatively agree with recent
optical conductivity experiments in detwinned BaFe$_{2}$As$_{2}$.
Experimental optical studies of compounds that display nematic order
without magnetic order, such as FeSe, would be desirable to further
elucidate this unavoidable entanglement between scattering rate and
plasma frequency anisotropies in these materials. 
\begin{acknowledgments}
We thank fruitful discussions with J. Chu, A. Chubukov, L. Degiorgi,
I. Fisher, D. Maslov, and S. Syzranov. M. S. acknowledges the support
from the Humboldt Foundation. J.S. acknowledges the support from Deutsche
Forschungsgemeinschaft (DFG) through the Priority Program SPP 1458
``Hochtemperatur-Supraleitung in Eisenpniktiden'' (project-no. SCHM
1031/5-1). R. M. F. is supported by the U.S. Department of Energy,
Office of Science, Basic Energy Sciences, under Award No. DE-SC0012336. 
\end{acknowledgments}

 \bibliographystyle{apsrev4-1}
\bibliography{bibliographyMod}

\section{Supplementary Material: ``Origin of the resistivity anisotropy in
the iron pnictides: scattering rate or plasma frequency?''}

\subsection{Form of the Geometric Coefficient}

\begin{figure}[h]
\includegraphics[width=0.3\textwidth]{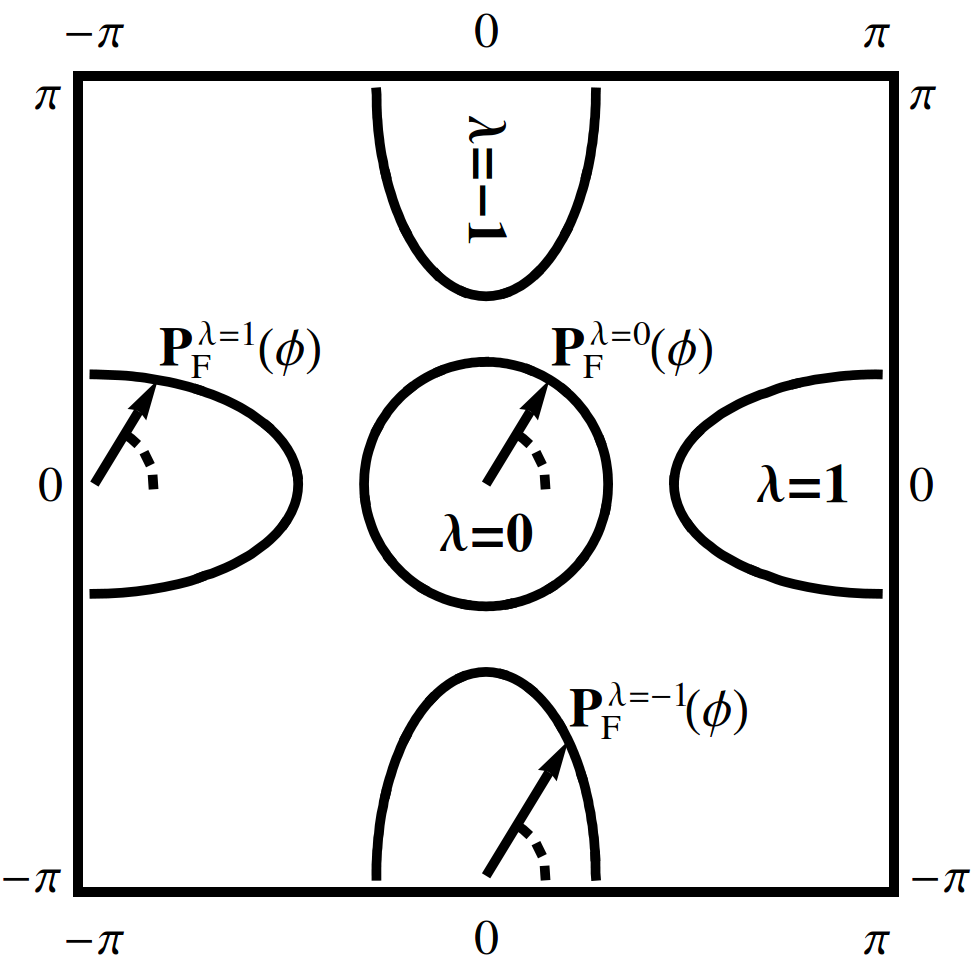}
\protect\caption{Illustration of the direction resolved parametrization of the Fermi
surface and the labeling of the bands involved.}

\label{FigDirectionParametrization} 
\end{figure}

The geometric coefficient $C_{\mathrm{eff}}^{\beta}$ appearing in
Eq. (4) of the main text is determined using direction-resolved Fermi
momentum $\vec{p}_{F}(\phi)$ and Fermi velocity $\vec{v}_{F}(\phi)$~\cite{SchuettPRL2015}.
Consequently this yields a direction-resolved density of states $N(\phi)=\vec{p}_{F}^{2}(\phi)/(2\pi\vec{v}_{F}(\phi)\vec{p}_{F}(\phi))$,
and the density of states of a single band $\nu_{F}^{\beta}=\int_{0}^{2\pi}N^{\beta}(\phi)\mathrm{d}\phi/(2\pi)$.
Using this notation, we can express the direction-average of an arbitrary
function $f\left(\phi\right)$ according to $\nu_{F}\braket{f(\phi)}_{\phi}=\int_{0}^{2\pi}N(\phi)f(\phi)\mathrm{d}\phi/(2\pi)$.
The angle $\phi$ here is always measured with respect to the x-axis
in each pocket, as shown in Fig~\ref{FigDirectionParametrization}.

Two bands that satisfy the constraint $|\beta+\beta'|=1$ specify
which of the two electron bands is interacting with the hole band
in the Brillouin center. The anisotropy arises from the hot spots,
the points on the Fermi surface which upon translation by either $\mathbf{Q}_{X}=\left(\pi,0\right)$
or $\mathbf{Q}_{Y}=\left(0,\pi\right)$ connect the two involved bands,
i.e. $\epsilon_{\mathbf{k}}^{\beta}=\epsilon_{\mathbf{k}+\mathbf{Q}_{j}}^{\beta'}$.
These hot spot are characterized by the hot spot angles $\phi_{\mathrm{HS}(\beta\beta')}$.
In terms of these quantities, the geometric coefficient is given by: 

\begin{align}
C_{\mathrm{eff}}^{\beta} & =\sum_{\substack{\beta'\\
|\beta+\beta'|=1
}
}\left(\beta+\beta'\right)\frac{N^{\beta}(\phi_{\mathrm{HS}(\beta\beta')})N^{\beta'}(\phi_{\mathrm{HS}(\beta\beta')})}{\left|\frac{\partial\vec{p}_{F}^{\beta}}{\partial\phi_{\mathrm{HS}(\beta\beta')}}\times\frac{\partial\vec{p}_{F}^{\beta'}}{\partial\phi_{\mathrm{HS}(\beta\beta')}}\right|}\nonumber \\
 & \times\frac{\left[v_{x,F}^{\beta}(\phi_{\mathrm{HS}(\beta\beta')})-v_{x,F}^{\beta'}(\phi_{\mathrm{HS}(\beta\beta')})\right]^{2}}{2\pi\nu_{F}^{\beta}\nu_{F}^{\beta'}\braket{(v_{x,F}^{\beta})^{2}}_{\phi}}
\end{align}
For the coefficient of the hole pocket $\beta=0$, the summation includes
the other two electron bands. However since the objects above obey
the $C_{4}$ symmetry of the system ($f^{\beta}(\phi+\pi/2)=f^{-\beta}(\phi)$
and $\vec{g}^{\beta}(\phi+\pi/2)=-i\sigma_{y}\vec{g}^{-\beta}(\phi)$
for any $C_{4}$ symmetric scalar function $f$ or vector function
$\vec{g}$), the contribution can be re-expressed in the familiar
form that is proportional to $(v_{x}^{\beta}-v_{x}^{\beta'})^{2}-(v_{y}^{\beta}-v_{y}^{\beta'})^{2}$.

\subsection{Estimative of the impurity scattering rate}

We estimated the impurity scattering rate from the residual resistivity
$\rho_{0}=\tau_{0}^{2}m/(ne^{2})$. For simplicity, we consider the
system as a cylindrical Fermi surface, such that the density can be
expressed according to $n=k_{F}^{2}/(2\pi c)$. Consequently the scattering
rate is found to be: 
\begin{equation}
\tau_{0}^{-1}h=\rho_{0}e^{2}hk_{F}^{2}/(2\pi mc)=\rho_{0}\hbar\bar{k}_{F}^{2}e^2/(mV)
\end{equation}
where in the last step we measured the Fermi momentum in units of
the inverse lattice constant $a$ and the unit cell volume $V=a^{2}c$
has been used. Using typical values for the 122 pnictides~\cite{JohnstonAP2010}:
we have $\bar{k}_{F}\approx\pi/4$, $a=4\,\textrm{\AA}$, $c=6.5\,\textrm{\AA}$,
$\rho_{0}\approx0.3\,\mathrm{m}\Omega\cdot\mathrm{cm}=3\times10^{-6}\Omega\cdot\mathrm{m}$,
yielding: 
\begin{equation}
\tau_{0}^{-1}h\approx330\,\mathrm{meV}
\end{equation}

\end{document}